\documentstyle[12pt,epsf]{article}
\def\lsim{<\kern-2.5ex\lower0.85ex\hbox{$\sim$}\ }
\def\rsim{>\kern-2.5ex\lower0.85ex\hbox{$\sim$}\ }
\begin{document}

\baselineskip 17pt

\centerline{\large\bf Measuring the Phase Velocity of Light}
\centerline{\large\bf in a Magnetic Field with the PVLAS
detector.}

\vspace{.25in}

 \centerline{A.C. Melissinos} \centerline{Department
of Physics, University of Rochester, Rochester, NY}
\centerline{\today}

\vspace{.50in} It is well known that the phase velocity of light
propagating in a transverse magnetic field is modified from its
vacuum value [1,2]. To lowest order in $\alpha$ one finds (in SI
units)

\begin{equation}
\frac{v}{c} = 1 - a \frac{2\alpha^2(\hbar c)^3}{45(mc^2)^4} \
\frac{B^2}{\mu_0} \sin^2\theta_B
\end{equation}
\noindent Here

$$\begin{array}{cllll}

a = 4 & \qquad & \qquad & \qquad {\rm If\ the\ polarization\ ({\it \vec{E}}-vector)\ of\ the}\\
& \qquad & \qquad & \qquad {\rm incident\ light\ is\ in\ the\ plane}\\
& \qquad & \qquad & \qquad {\rm defined\ by\ the\ wave\ vector}\ {\it \vec{k}}\\
& \qquad & \qquad & \qquad {\rm and\ the\ field\ {\it \vec{B}}}\\
\\
a = 7 & \qquad & \qquad & \qquad {\rm If\ the\ polarization\ is\
perpendicular}\\
& \qquad & \qquad & \qquad {\rm to\ that\ plane}\\
\\
\theta_B & \qquad & \qquad & \qquad{\rm Is\ the\ angle\ between\
{\it \vec{k}}\ and\ {\it \vec{B}}}

\end{array}$$

\noindent Numerically and for $\theta_B = 90^\circ$,

\begin{equation}
\frac{v}{c} = 1 - a (1.3 \times 10^{-24})\ \left(\frac{B}{1\ {\rm
T}}\right)^2
\end{equation}

Attempts to measure this effect have been centered on the
ellipticity induced on linearly polarized light incident at
45$^\circ$ to the field direction (and $\theta_B = 90^\circ$)
[3,4,5].  In a recent paper Boer and van Holten [6] propose a
direct measurement of the phase velocity using the large
gravitational interferometers, such as LIGO [7].  In their scheme
one needs to modulate the magnetic field which forces the signal
to low frequencies where the noise is excessive (seismic noise).

On the other had a detector such as PVLAS [5] could be used to
make this measurement as discussed below.  The magnetic field is
kept constant (contrary to the usual rotating configuration
employed in PVLAS) but instead the linear polarization is rotated
at a frequency $f_R <1/(2 \pi\tau_s)$, where $\tau_s$ is the
storage time in the cavity.  As the polarization changes from
$\parallel$ to $\perp$ the phase velocity changes according to
eq.(2) from $a = 4$ to $a = 7$.  To maintain the cavity on
resonance the servo system will have to provide a correction
signal at $2f_R$.  This signal may be detectable above the noise
if long integration times are used.

To estimate the signal we use the PVLAS parameters [5]

$$
\begin{array}{lllll}
B & = & 7 & {\rm T}\\
\ell & = & 1 & {\rm m}\\
F & = & 2 \times 10^5 & ({\rm finesse})\\
\lambda & = & 532 & {\rm nm}
\end{array}
$$

\noindent to find

$$\begin{array}{llll}
\Delta (v/c) & = & 2 \times 10^{-22}\\
\Delta\ell/\ell & = & 4 \times 10^{-17}\ {\rm m}\\
\Delta\phi & = & 4 \pi \Delta \ell F/\lambda \simeq 10^{-9}\\
\tau_s & = & F\ell/\pi c \simeq 2 \times 10^{-4}\ {\rm s}
\end{array}
$$

The resulting phase shift is indeed small but comparable to the
ellipticities that PVLAS can measure.  However here the
polarizastion can be rotated at $f_R \rsim 1\ {\rm kHz}$ and if
this is done by an E/O device there is no mechanical noise at
$f_R$.  Finally, PVLAS has the advantage of being currently
operational.  We realize that rotating the polarization poses
problems for the Pound-Drever [8] scheme of locking the NPRO laser
onto the cavity, but hope that these can be overcome.

\vspace{.25in}

\noindent{\bf References}

\begin{enumerate}

\item W. Heisenberg and H. Euler, Z. Phys. \underline{98} (1936) 714; J.
Schwinger, Phys. Rev. \underline{82} (1951) 664.

\item S.L. Adler, Ann. Phys. \underline{67} (1971) 599.

\item E. Iacopini and E. Zavattini, Phys. Lett. \underline{B85} (1979)
151.

\item R. Cameron {\it et al.} Phys. Rev. \underline{D47}, (1993) 3707.

\item G. Cantatore {\it et al.}, Phys. Lett. \underline{B265}
(1991) 418;  E. Zavattini {\it et al.} \lq\lq The PVLAS
Collaboration" in \lq\lq Quantum Electrodynamics and Physics of
the Vacuum", AIP Conference Proceedings \underline{564}, New York,
NY (2001), G. Cantatore Editor.

\item D. Boer and J.-W. van Holten arXiv: hep-ph/0204207v1.

\item A. Abramovici {\it et al.}, Science \underline{256} (1992)
325;\ http://www.ligo.caltech.edu

\item R.P. Drever {\it et al.}, Appl. Phys. \underline{B31}
(1983) 97.

\end{enumerate}

\end{document}